\documentstyle[11pt,ysc,twoside,epsf]{article}

\def\edcomment#1{\iffalse\marginpar{\raggedright\sl#1\/}\else\relax\fi}

\reversemarginpar

\begin{document}

\title{Diffuse Interstellar Band at 5850 as a Member of 5797 Spectroscopic Family}

\author{Katarzyna Bryndal}

\affil{ Jan D\l ugosz Academy, Institute of Physics,
Cz\c{e}stochowa, Al. Armi Krajowej 13/15, 42-200 Cz\c{e}stochowa,
Poland.}

\author{Bogdan Wszo\l ek}

\affil{ Jan D\l ugosz Academy, Institute of Physics,
Cz\c{e}stochowa, Al. Armi Krajowej 13/15, 42-200 Cz\c{e}stochowa,
Poland. \\ E-mail:bogdan{@}ajd.czest.pl }

\begin{abstract}

The carriers of diffuse interstellar bands are still mysterious
species. There exist many arguments that diffuse bands at 5797 and
5850 angstroms have the same carrier. Using high-resolution spectra
of few dozens of reddened stars we have searched mutual correlation
between intensities of considered bands. Results of our analysis
indicate that 5797 and 5850 really tend to have the same carrier.

\end{abstract}

\section{Introduction}

Diffuse interstellar bands (DIBs) are mysterious absorption
structures of interstellar origin observed in spectra of reddened
stars of early spectral types. The detection of the first two DIBs
was described by Heger in 1922. These structures, as we know them
today, are scattered within the whole region of visible light and in
the near infrared (see e.g. Herbig 1995). Individual DIBs differ
between themselves in intensity, line width and profile shapes.

There are many arguments that 5797 and 5850 DIBs have the same
carrier. Almost all procedures isolating DIBs' families indicate
that 5797 and 5850 bands tend to belong to the same family (e.g.
Chlewicki et al. 1986, Josafatson and Snow 1987, Kre\l owski and
Walker 1987, Wszo\l ek and God\l owski 2003).

Wszo\l ek and God\l owski (2003) additionally confirmed the status
of 5850 band as a member of 5797 spectroscopic family, that means
that both these DIBs have probably the same carrier. They composed
sequences of spectrograms, of different target stars, in such a way
that equivalent width (EW) of 5797 was substantially changing from
one case to the other. They realized that the depth of 5850 band
regularly follows the depth of 5797 band.

To check how good candidate to 5797 spectroscopic family is the DIB
at 5850 angstroms, we carried out measurements of EWs for both DIBs
in spectra of 74 target stars. We also carried out the analysis of
mutual correlation between them.

\section{ Observational Data }

All the spectra analyzed for the purpose of this contribution were
taken from the archives of Professor Jacek Kre\l owski (Astronomical
Centre, Nicolaus Copernicus University, Toru\'{n}, Poland). We used
spectra taken with the Canada-France-Hawaii Telescope and spectra
acquired at the McDonald Observatory with an echelle spectrograph
fed with the 2.1-m telescope. Considered DIBs were present always in
the same order of the spectra. Spectra are of quite good quality, as
far as resolution and signal/noise are concerned.

\section{ Correlation Analysis }

Using spectra of 74 target stars we have measured EWs for DIBs at
5797 and 5850 angstroms. That was done with the help of software
package REWIA v. 1.4, written by Jerzy Borkowski (Nicolaus
Copernicus Astronomical Center, Polish Academy of Sciences,
Toru\'{n}, Poland). We calculated correlation coefficient for
measured values. We drew correlation diagram, which is exposed in
figure 1. On the figure one may notice small dispersion of points
around the straight line. It is most probably caused by errors in
observation and in data reduction. Correlation coefficient r=0.9 was
counted and it is very high. In practice, there exist no other pair
of DIBs with such good correlation as 5797 and 5850 ones.

\begin{figure}
\plotone{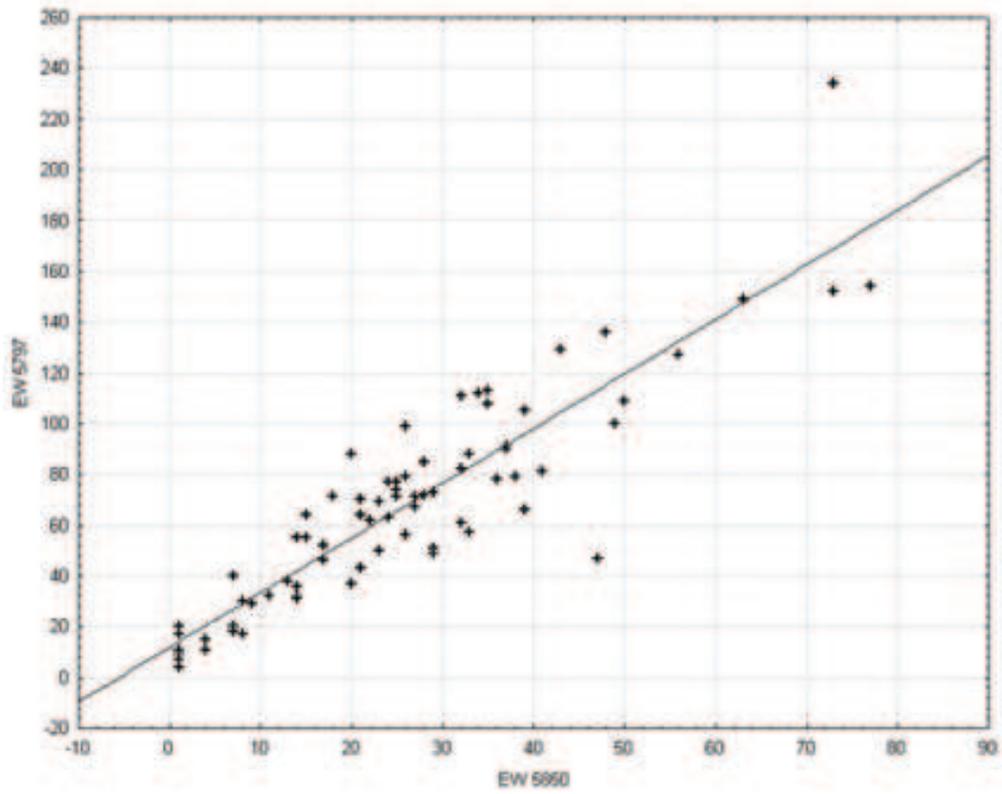} \caption{Correlation diagram between EW(5850)
and EW(5797) for 74 target stars. EWs are given in miliangstroms}
\end{figure}

\section{Concluding Remarks}

5850 band lies in the spectra very close to 5797 DIB. Although 5850
is substantially weaker than 5797, it is still a relatively strong
DIB. That means that for very reddened stars EW(5850) measurement
errors are not too big. Furthermore, 5850 morphologically resembles
very much the 5797 band. All groupings of DIBs into families, known
from literature, show that 5850 and 5797 belong to the same
morphological family. Correlation diagram for large number of target
stars, exposed in the figure 1, and counted coefficient of
correlation strengthen very much the hypothesis that 5797 and 5850
DIBs have the same carrier.

\section{Acknowledgements}

We would like to thank Professor Jacek Kre\l owski for giving us an access to his data archives.

\begin {references}

\reference Chlewicki G., van der Zwet G.P., van Ijzendoorn L.J.,
Greenberg J.M.: 1986, ApJ., 305, p.455;

\reference Heger M.L.: 1922, Lick. Obs. Bull., 10, p.146;

\reference Herbig G.H.: 1995, ARA\&A, 33, p.19;

\reference Josafatson K., Snow T.P.: 1987, ApJ., 319, p.436;

\reference Kre\l owski J., Walker G.A.H.: 1987, ApJ., 312, p.860;

\reference Wszo\l ek B., God\l owski W.: 2003, MNRAS, 338, p.990;

\end {references}

\end{document}